\input epsf
\input rotate
\documentstyle[bolditalic,prb,aps,12pt]{revtex}

\begin{document}

\title{Two-fluid magnetic island dynamics in slab geometry:\\
II - Islands interacting with resistive walls or static\\  external resonant magnetic perturbations}
\author{~\\Richard Fitzpatrick\thanks{rfitzp@farside.ph.utexas.edu}
and Fran\c{c}ois L.\ Waelbroeck\\
{\em Institute for Fusion Studies}\\
{\em Department of Physics}\\
{\em University of Texas at Austin}\\
{\em Austin, TX 78712}\\
~\\}
\maketitle
\begin{abstract} 
The dynamics of a propagating magnetic island
interacting with a resistive wall or a static external resonant magnetic perturbation
is investigated
using two-fluid, drift-MHD (magnetohydrodynamical) theory in {\em slab geometry}. In both cases,  the island equation of motion is found to take exactly the same form
as that predicted by single-fluid MHD theory.
{\em Three} separate ion polarization terms are found in the Rutherford
island width evolution equation. The first is the drift-MHD polarization term for an isolated
island, and is completely unaffected by the interaction with a
wall or  magnetic perturbation. Next, there is the
polarization term due to interaction with a  wall or magnetic
perturbation which is predicted by {\em single-fluid} MHD theory. This term is
always {\em destabilizing}. Finally, there is a hybrid of the other
two polarization terms. The
sign of this term depends on many factors. However, under normal
circumstances, it is stabilizing if the unperturbed island  propagates in the
{\em ion diamagnetic direction} (in the lab.\ frame), and {\em destabilizing}\/ if it
propagates in the {\em electron diamagnetic
direction}.
\end{abstract}

\section{Introduction}
Tearing modes are magnetohydrodynamical (MHD) instabilities
which often limit fusion plasma performance in magnetic 
confinement devices relying on nested toroidal magnetic
 flux-surfaces.\cite{ros} As the name suggests, ``tearing'' modes tear and reconnect magnetic
field-lines, in the process converting  nested toroidal flux-surfaces into helical magnetic islands.
Such islands degrade plasma confinement because
heat and particles are able to travel radially from one side of an island
to another by flowing along magnetic field-lines, which is a
relatively fast process, instead of having to diffuse across
magnetic flux-surfaces, which is a relatively slow process.\cite{callen}

The interaction of rotating magnetic islands with resistive walls\cite{snipes,hender,zohm,nave,rf,gates,wf,guo,chapman} and
external resonant magnetic perturbations\cite{zohm,rf,morris,navratil,rf1} has been the subject of a great deal of research in the magnetic fusion community,
since such interactions can  have a highly deleterious effect on
plasma confinement. This paper focuses on the
{\em ion polarization} corrections to the Rutherford island width evolution equation\cite{ruth} which arise from the highly sheared ion flow
profiles generated around magnetic islands whose rotation frequencies
are shifted by interaction with either  resistive walls or
external magnetic perturbations. According to  {\em single-fluid} MHD (magnetohydrodynamical) theory,\cite{wf,rf1} such polarization corrections are
always {\em destabilizing}. The aim of this paper is to evaluate the
ion polarization corrections using {\em two-fluid},
drift-MHD theory, which is far  more relevant to present-day magnetic
confinement devices than single-fluid theory. This goal is achieved
by extending the analysis of the companion paper,\cite{island1} which
investigates the dynamics of an {\em isolated} magnetic island in slab geometry using
two-fluid, drift-MHD theory. For the sake of simplicity, we shall restrict our
investigation to {\em slab geometry}.

\section{Reduced equations}
\subsection{Basic equations}
Standard right-handed Cartesian coordinates ($x$, $y$, $z$) are adopted.
Consider a quasi-neutral plasma with singly-charged ions of mass
$m_i$. The 
ion/electron number density $n_0$ is assumed to be {\em uniform} and {\em constant}.
Suppose that $T_i = \tau\,T_e$, where $T_{i,e}$ is the ion/electron
temperature, and $\tau$ is {\em uniform} and {\em constant}. 
Let there be no
variation of quantities in  the $z$-direction: {\em i.e.}, $\partial/\partial z\equiv 0$. 
Finally, let all lengths be normalized to some convenient scale length $a$,
all magnetic field-strengths to some convenient scale field-strength $B_a$,
and all times to $a/V_a$, where $V_a = B_a/\sqrt{\mu_0\,
n_0\,m_i}$.

We can write ${\bfm B} = 
 \nabla\psi\times \hat{\bfm z}+ (B_0+b_z)\,\hat{\bfm z}$, and $P= P_0
 -B_0\,b_z + O(1)$, where ${\bfm B}$ is the magnetic field, and $P$ the
 total plasma pressure. Here, we are assuming that $B_0 \gg P_0\gg 1$,
 with $\psi$ and $b_z$ both $O(1)$.\cite{island1} Let, $\beta = \Gamma\,P_0/B_0^{\,2}$ be ($\Gamma$ times) the plasma beta calculated with the ``guide-field'', $B_0$, where $\Gamma=5/3$ is the plasma ratio of specific heats.
 Note that the above ordering scheme does not
constrain $\beta$ to be either much less than or much greater than unity.

We adopt the reduced, 2-D, two-fluid, drift-MHD equations derived in the companion
paper:\cite{island1}
\begin{eqnarray}
\frac{\partial\psi}{\partial t}&=& [\phi - d_\beta\,Z,\psi]
+ \eta\,(J-J_0) - \frac{\mu_e\,d_\beta\,(1+\tau)}{c_\beta}\,
\nabla^2[V_z + (d_\beta/c_\beta)\,J],\label{efa}\\[0.5ex]
\frac{\partial Z}{\partial t} &=& [\phi,Z] + c_\beta\,[V_z+
(d_\beta/c_\beta)\,J,\psi] + c_\beta^{\,2}\,D \,Y + \mu_e\,d_\beta\,\nabla^2(U-d_\beta\,Y),\\[0.5ex]
\frac{\partial U}{\partial t} &=& [\phi,U]
- \frac{d_\beta\,\tau}{2} \left\{\nabla^2[\phi,Z]+[U,Z]+[Y,\phi]\right\}
+ [J,\psi]+\mu_i\,\nabla^2(U + d_\beta\,\tau\,Y) \nonumber\\[0.5ex]&&+ 
\mu_e\,\nabla^2(U-d_\beta\,Y),\\[0.5ex]
\frac{\partial V_z}{\partial t} &=& [\phi,V_z]+c_\beta\,[Z,\psi]
+\mu_i\,\nabla^2 V_z + \mu_e \,\nabla^2[V_z+ (d_\beta/c_\beta)\,J],\label{efd}
\end{eqnarray}
where $D= \eta\,(1-(3/2)\,[\tau/(1+\tau)])+\kappa/\beta$, $U=\nabla^2\phi$, $J=\nabla^2\psi$, and $Y=\nabla^2 Z$.
Here, $c_\beta=
\sqrt{\beta/(1+\beta)}$, $d_\beta = c_\beta\,d_i/\sqrt{1+\tau}$, 
$Z=b_z/c_\beta\,\sqrt{1+\tau}$,  $d_i = (m_i/n_0\,e^2\,\mu_0)^{1/2}/a$,
and $[A,B] = \nabla A\times \nabla B\!\cdot\!\hat{\bfm z}$. The {\em guiding-center}
velocity is written: ${\bfm V} = \nabla\phi\times\hat{\bfm z} + \sqrt{1+\tau}\,V_z$. Furthermore, $\eta$ is the (uniform) plasma resistivity, $\mu_{i\,e}$
the (uniform) ion/electron viscosity,  $\kappa$ the (uniform)
plasma thermal conductivity, and $J_0(x)$  (minus) the inductively
maintained, equilibrium plasma current in the $z$-direction.
The above equations contain both electron and ion diamagnetic effects,
including the contribution of the anisotropic ion gyroviscous tensor, but
neglect electron inertia. Our equations are ``reduced" in the sense that they
do not
contain the compressible Alfv\'{e}n wave. However, they do contain the
shear-Alfv\'{e}n wave, the magnetoacoustic wave, the whistler wave,
and the kinetic-Alfv\'{e}n wave.

\subsection{Plasma equilibrium}
The plasma equilibrium satisfies $\partial/\partial y\equiv 0$. Suppose that
the plasma is bounded by rigid walls at $x=\pm x_w$, and that the region
beyond the walls is a vacuum.
The equilibrium
magnetic flux is written $\psi^{(0)}(x)$, where $\psi^{(0)}(-x)=\psi^{(0)}(x)$, and $d^2\psi^{(0)}(x)/dx^2=J_0(x)$.  The scale
magnetic field-strength, $B_a$, is chosen such that $\psi^{(0)}(x)
\rightarrow -x^2/2$ as $|x|\rightarrow 0$.
The equilibrium value of the field $Z$ takes the form $Z^{(0)}(x)
=-[V_{\ast\,y}^{(0)}/d_\beta\,(1+\tau)]\,x$, where $V_{\ast\,y}^{(0)}$ is the (uniform)
total diamagnetic velocity in the $y$-direction. 
The equilibrium value of the guiding-center stream-function is written $\phi^{(0)}(x)
= -V_{EB\,y}^{(0)}\,x$, where $V_{EB\,y}^{(0)}$ is the (uniform)
equilibrium ${\bfm E}\times{\bfm B}$ velocity in the $y$-direction.
Finally, the equilibrium value of the field $V_z$ is simply $V_z^{(0)}=0$.

\subsection{Asymptotic matching}
Consider a tearing perturbation which is periodic in the $y$-direction
with periodicity length $l$.
According to conventional analysis, the plasma is conveniently split into two regions.\cite{fkr} The ``outer region''
comprises most of the plasma, and is governed by the equations of
linearized, ideal-MHD. On the other hand, the ``inner region'' is localized in the
vicinity of the magnetic resonance $x=0$ (where $B_y^{(0)}=0$). Non-linear, dissipative, and drift-MHD effects all become
important in the inner region.

In the outer region, we can write $\psi(x,y,t)=\psi^{(0)}(x)+
\psi^{(1)}(x,t)\,\exp({\rm i}\,k\,y)$, where $k=2\pi/l$ and
$|\psi^{(1)}|\ll|\psi^{(0)}|$. Linearized ideal-MHD yields
$[\psi^{(1)},J^{(0)}]+[\psi^{(0)},J^{(1)}] =0$, where $J=\nabla^2\psi$.
It follows that
\begin{equation}\label{tear}
\left(\frac{\partial^2}{\partial x^2}-k^2\right)\psi^{(1)}
-\left(\frac{d^3\psi^{(0)}/d x^3}{d\psi^{(0)}/dx}\right)\psi^{(1)}=0.
\end{equation}
The solution to the above equation must be asymptotically matched to the
full, non-linear, dissipative, drift-MHD solution in the inner region.

\section{Interaction with a resistive wall}
\subsection{Introduction}\label{rint}
Suppose that the walls bounding the plasma at $x=\pm x_w$ are  thin and
resistive, with time-constant $\tau_w$. We can define the perfect-wall
tearing eigenfunction, $\psi_{pw}(x)$, as the continuous even (in $x$) solution to Eq.~(\ref{tear})
which satisfies $\psi_{pw}(0)=1$, and $\psi_{pw}(\pm x_w)=0$. Likewise,
the no-wall tearing eigenfunction, $\psi_{nw}(x)$, is the continuous even solution to Eq.~(\ref{tear})
which satisfies $\psi_{pw}(0)=1$, and $\psi_{pw}(\pm\infty)=0$.
In general, both $\psi_{pw}(x)$, and $\psi_{nw}(x)$ have gradient
discontinuities at $x=0$. The quantity ${\mit\Delta}_{pw}=[d\psi_{pw}/dx]_{0-}^{0+}$ is the conventional tearing stability index\cite{fkr} in the
presence of a perfectly conducting wall ({\em i.e.}, $\tau_w\rightarrow\infty$), whereas ${\mit\Delta}_{nw}=[d\psi_{nw}/dx]_{0-}^{0+}>{\mit\Delta}_{pw}$ is the tearing stability index in the
presence of no wall ({\em i.e.}, $\tau_w\rightarrow 0$). 
Finally, the wall eigenfunction, $\psi_w(x)$, is defined as the continuous even solution to Eq.~(\ref{tear})
which satisfies $\psi_w(0)=0$, $\psi_w(\pm x_w)=1$, and $\psi_w(\pm\infty)=0$. This eigenfunction
has  additional gradient discontinuities at $x=\pm x_w$. The
wall stability index, ${\mit\Delta}_w<0$,  is defined ${\mit\Delta}_w=[d\psi_w/dx]_{x_w-}^{x_w+}$.

According to standard analysis,\cite{rf} the effective tearing stability
index, ${\mit\Delta}' = [d\ln\psi/dx]_{0-}^{0+}$, in the presence of a resistive wall is written
\begin{equation}\label{del}
{\mit\Delta}' = \frac{V^2\,{\mit\Delta}_{pw}+V_w^{\,2}\,{\mit\Delta}_{nw}}{V^2+V_w^{\,2}},
\end{equation}
where $V$ is the phase-velocity of the tearing mode in the lab.\ frame, and
$V_w = (-{\mit\Delta}_w)/(k\,\tau_w)$. Also, the net $y$-directed electromagnetic 
force acting on the inner region takes the form
\begin{equation}\label{force}
f_y = -\frac{k}{2}\,({\mit\Delta}_{nw}-{\mit\Delta}_{pw})\,
\frac{V\,V_w}{V^2+V_w^{\,2}}\,{\mit\Psi}^2,
\end{equation}
where ${\mit\Psi}(t)=|\psi^{(1)}(0,t)|$ is the  reconnected magnetic flux,
which is assumed to have a very weak time dependence.

\subsection{Island geometry}
In the inner region, we can write
\begin{equation}
\psi(x,\theta,t) = -\frac{x^2}{2} + {\mit\Psi}(t)\,\cos\theta,
\end{equation}
where $\theta = k\,y$. As is well-known, the above expression for $\psi$ describes
a constant-$\psi$ magnetic island of full-width (in the $x$-direction)
$W=4\,w$, where $w=\sqrt{\mit\Psi}$. 
The region inside the magnetic separatrix corresponds to $\psi>-{\mit\Psi}$, 
whereas the region outside the separatrix corresponds to $\psi\leq -{\mit\Psi}$.
It is convenient to work in the {\em island rest frame}, in which $\partial/\partial t\simeq 0$.

It is helpful to define a flux-surface average operator:
\begin{equation}
\langle f(s,\psi,\theta) \rangle= \oint \frac{f(s,\psi,\theta)}{|x|}\,\frac{d\theta}{2\pi}
\end{equation}
for $\psi\leq -{\mit\Psi}$, and
\begin{equation}
\langle f(s,\psi,\theta) \rangle = \int_{-\theta_0}^{\theta_0}
\frac{f(s,\psi,\theta)+f(-s,\psi,\theta)}{2\,|x|}\,\frac{d\theta}{2\pi}
\end{equation}
for $\psi> -{\mit\Psi}$. Here, $s={\rm sgn}(x)$, and $x(s,\psi,\theta_0)=0$
(with $\pi>\theta_0>0$).
The most important property of this operator is that
$\langle [A,\psi]\rangle \equiv 0$,
for any field $A(s,\psi,\theta)$. 

\subsection{Ordering scheme}
In the inner region, we adopt the following ordering of terms appearing in Eqs.~(\ref{efa})--(\ref{efd}):
$\psi = \psi^{(0)}$, $\phi=\phi^{(1)}(s,\psi) + \phi^{(3)}(s,\psi,\theta)$,
$Z= Z^{(1)}(s,\psi)+Z^{(3)}(s,\psi,\theta)$, $V_z=V_z^{(2)}(s,\psi,\theta)$, $\delta J = 
\delta J^{(2)}(s,\psi,\theta)$. Moreover, $\nabla=\nabla^{(0)}$, $\tau=\tau^{(0)}$,
$c_\beta = c_\beta^{(0)}$, $d_\beta =d_\beta^{(0)}$, $\mu_{i,e}=
\mu_{i,e}^{(2)}$, $\kappa=\kappa^{(2)}$, $\eta =\eta^{(2)}$,
and $d{\mit\Psi}/dt=d{\mit\Psi}^{(4)}/dt$. 
Here, the superscript $^{(i)}$ indicated an $i$th order quantity.
This ordering, which is completely self-consistent,  implies weak ({\em i.e.},
strongly sub-Alfv\'{e}nic and sub-magnetoacoustic) diamagnetic  flows, and very long ({\em i.e.}, very much
longer than the Alfv\'{e}n time) transport evolution time-scales.

To lowest and next lowest orders, Eqs.~(\ref{efa})--(\ref{efd}) yield:
\begin{eqnarray}
\frac{d{\mit\Psi}^{(4)}}{dt}\,\cos\theta&=& [\phi^{(3)} - d_\beta\,Z^{(3)},\psi]
+ \eta^{(2)}\,\delta J^{(2)} - \frac{\mu_e^{(2)}\,d_\beta\,(1+\tau)}{c_\beta}\,
\nabla^2[V_z^{(2)} + (d_\beta/c_\beta)\,\delta J^{(2)}]\label{eva},\\[0.5ex]
0&=& c_\beta\,[V_z^{(2)}+
(d_\beta/c_\beta)\,\delta J^{(2)},\psi] +c_\beta^{\,2}\,D^{(2)}\,Y^{(1)} + \mu_e^{(2)}\,d_\beta\,\nabla^2(U^{(1)}-d_\beta\,Y^{(1)}),\label{evb}\\[0.5ex]
0 &=&- M^{(1)}\,[U^{(1)},\psi]
- \frac{d_\beta\,\tau}{2} \left\{L^{(1)}\,[U^{(1)},\psi]+M^{(1)}\,[Y^{(1)},\psi]\right\}
+ [\delta J^{(2)},\psi]\nonumber\\[0.5ex]&&+\mu_i^{(2)}\,\nabla^2(U^{(1)} + d_\beta\,\tau\,Y^{(1)})+ 
\mu_e^{(2)}\,\nabla^2(U^{(1)}-d_\beta\,Y^{(1)}),\label{evc}\\[0.5ex]
0&=&- M^{(1)}\,[V_z^{(2)},\psi]+c_\beta\,[Z^{(3)},\psi]
+\mu_i^{(2)}\,\nabla^2 V_z^{(2)} + \mu_e^{(2)} \,\nabla^2[V_z^{(2)}+ (d_\beta/c_\beta)\,\delta J^{(2)}]\label{evd}
\end{eqnarray}
in the inner region, where $\delta J^{(2)} = J+1$, $Y^{(1)}=\nabla^2 Z^{(1)}$, $U^{(1)}=\nabla^2\phi^{(1)}$,
$M^{(1)}(s,\psi)=d\phi^{(1)}/d\psi$, and $L^{(1)}(s,\psi)=dZ^{(1)}/d\psi$.
Here, we have neglected the superscripts on zeroth order quantities,
for the sake of clarity. 
In the following, we shall neglect all superscripts,
except for those on $\phi^{(3)}$ and $Z^{(3)}$, for ease of notation.

\subsection{Determination of flow profiles}\label{flow}
Flux surface averaging Eqs.~(\ref{evb}) and (\ref{evc}), we obtain 
\begin{equation}\label{ex}
\langle \nabla^2 U\rangle + \frac{d_\beta\,(\mu_i\,\tau-\mu_e)}
{(\mu_i+\mu_e)}\,\langle\nabla^2 Y\rangle = 0,
\end{equation}
and
\begin{equation}\label{ey}
\delta^2\,w^2\,\langle \nabla^2 Y\rangle -\langle Y\rangle =0,
\end{equation}
where
\begin{equation}
\delta=\frac{d_i}{w\,\sqrt{D}}\,\sqrt{\frac{\mu_i\,\mu_e}{\mu_i+
\mu_e}}.
\end{equation}
In the following, we shall assume that $\delta\ll 1$.

 Now, we can
write $\nabla^2\simeq\partial^2/\partial x^2$, provided that  the island is ``thin'' ({\em i.e.}, $w\ll l$). It follows that
\begin{equation}\label{edd}
M(s,\psi) =  - \frac{d_\beta\,(\mu_i\,\tau-\mu_e)}
{(\mu_i+\mu_e)}\,L(s,\psi) + F(s,\psi),
\end{equation}
where
\begin{equation}\label{ez}
\frac{d}{d\psi}\!\left[\frac{d}{d\psi}\!\left(\delta^2\,w^2\,\langle x^4\rangle\,\frac{d L}{d
\psi}\right)-\langle x^2\rangle\,L\right] =0,
\end{equation}
and
\begin{equation}\label{ezz}
\frac{d^2}{d\psi^2}\!\left(\langle x^4\rangle\,\frac{d F}{d
\psi}\right)=0.
\end{equation}

Note that $L(s,\psi)$ and $F(s,\psi)$ are  {\em odd} functions of
$x$. We immediately conclude that $L(s,\psi)$ and $F(s,\psi)$ are both
{\em zero} inside the island separatrix (since it is impossible to have a non-zero
odd flux-surface function in this region). The function $L(s,\psi)$ satisfies
the additional boundary condition  $x\,L\rightarrow V^{(0)}_{\ast\,y}/
d_\beta\,(1+\tau)$ as $|x|/w\rightarrow\infty$. Here, we are assuming
that $w\ll x_w$. Moreover, the function $F(s,\psi)$ satisfies the
additional boundary condition $x\,F\rightarrow
(|x|/x_w)\,(V^{(0)}-V)$ as $|x|/w\rightarrow 0$, where $V^{(0)}$ is
the unperturbed island phase-velocity in the lab.\ frame. 

It is helpful to define the following quantities: $\hat{\psi}=-\psi/{\mit\Psi}$, $\langle\!\langle\cdots
\rangle\!\rangle= \langle\cdots\rangle\,w$, and $X=x/w$. 
The solutions to Eqs.~(\ref{ez}) and (\ref{ezz}), subject to the above
mentioned boundary conditions, are 
\begin{equation}\label{eleq}
L(s,\hat{\psi}) = \frac{s\,V_{\ast\,y}^{(0)}}{w\,d_\beta\,(1+\tau)}\,
\frac{1}{\langle\!\langle X^2\rangle\!\rangle},
\end{equation}
and
\begin{equation}\label{fdef}
F(s,\hat{\psi}) = \frac{s\,(V^{(0)}-V)}{x_w}\,
\left.\int_{1}^{\hat{\psi}}\frac{d\hat{\psi}}{\langle\!\langle
X^4\rangle\!\rangle}\right/ \int_{1}^{\infty}\frac{d\hat{\psi}}{\langle\!\langle
X^4\rangle\!\rangle},
\end{equation}
respectively.
Of course, both $L(s,\hat{\psi})$ and $F(s,\hat{\psi})$ are zero inside
the island separatrix ({\em i.e.}, $\hat{\psi}<1$). In writing Eq.~(\ref{eleq}),
we have neglected the thin boundary layer  (width, $\delta\,w$) which
resolves the apparent discontinuity in $L(s,\hat{\psi})$ across the
island separatrix. This
boundary layer, which need not be resolved in any of our
calculations, is described in the companion paper.\cite{island1}
Note that the function $L(s,\hat{\psi})$ corresponds to a velocity profile
which is {\em localized} in the vicinity of the island, whereas the
function $F(s,\hat{\psi})$ corresponds to a {\em non-localized} profile which extends over the
whole plasma.

\subsection{Force balance}\label{fbal}
The net electromagnetic force acting on the island region can be
written\cite{rf1}
\begin{equation}\label{h1}
f_y = -2\,k\,{\mit\Psi}\,\int_{{\mit\Psi}}^{-\infty}
\langle \delta J_s\,\sin\theta\rangle\,d\psi,
\end{equation}
where $\delta J_s$ is the component of $\delta J$ with the
symmetry of $\sin\theta$.
Now, it is easily demonstrated that 
\begin{equation}\label{h2}
\langle \delta J_s\,\sin\theta\rangle = \frac{1}{k\,{\mit\Psi}}
\,\langle x\,[\delta J_s, \psi]\rangle,
\end{equation}
so it follows from Eq.~(\ref{evc}) that
\begin{equation}
\langle \delta J_s\,\sin\theta\rangle = -\frac{(\mu_i+\mu_e)}{k\,{\mit\Psi}}
\frac{d}{d\psi}\!\left(\langle
x^5\rangle\,\frac{d^2 F}{d\psi^2}- 2\,\langle x^3\rangle
\,\frac{d F}{d\psi} - \langle x\rangle F\right).
\end{equation}
Hence,
\begin{eqnarray}
f_y &=&  2\,(\mu_i+\mu_e)\lim_{x/w\rightarrow \infty}\left(\langle
x^5\rangle\,\frac{d^2 F}{d\psi^2}- 2\,\langle x^3\rangle
\,\frac{d F}{d\psi} - \langle x\rangle F\right) \nonumber\\[0.5ex]
&=&2\,s\,(\mu_i+\mu_e)\lim_{x/w \rightarrow \infty}\left[x^2\,\frac{d}{dx}\!\left(
\frac{1}{x}\,\frac{d(x\,F)}{dx}\right)\right] .
\end{eqnarray}
Finally, Eq.~(\ref{fdef}) yields
\begin{equation}\label{force1}
f_y = -\frac{2\,(\mu_i+\mu_e)\,(V^{(0)}-V)}{x_w}.
\end{equation}

Equating Eqs.~(\ref{force}) and (\ref{force1}), we obtain
the island force balance equation:
\begin{equation}
\frac{2\,(\mu_i+\mu_e)\,(V^{(0)}-V)}{x_w}=\frac{k}{2}\,({\mit\Delta}_{nw}-{\mit\Delta}_{pw})\,
\frac{V\,V_w}{V^2+V_w^{\,2}}\,(W/4)^4.
\end{equation}
This equation describes the competition between the viscous restoring
force (left-hand side) and the electromagnetic wall drag (right-hand side) acting on the island, and
determines the island phase-velocity, $V$, as a function of
the island width, $W$. Note that the above force balance equation is
identical to that obtained from single-fluid MHD theory.\cite{rf}

\subsection{Determination of ion polarization correction}\label{polz}
It follows from Eqs.~(\ref{eva}), (\ref{evc}), and (\ref{evd}) that
\begin{equation}\label{e29}
\delta J_c = -\frac{1}{2}\!\left(X^2-\frac{\langle\!\langle X^2\rangle\!\rangle}{\langle\!\langle 1
\rangle\!\rangle}\right) \frac{d}{d\hat{\psi}}\!\left[M\,(M+d_\beta\,\tau\,L)\right]
+\eta^{-1}\,\frac{d{\mit\Psi}}{dt}\,\frac{\langle\!\langle\cos\theta\rangle\!\rangle}{\langle\!\langle
1\rangle\!\rangle},
\end{equation}
where $\delta J_c$ is the component of $\delta J$ with the symmetry of
$\cos\theta$. In writing the above expression, we have neglected any
boundary layers on the island separatrix, since these are either unimportant or
 need
not be resolved in our calculations (see Ref.~\onlinecite{island1}).
Now, making use of Eqs.~(\ref{edd}), (\ref{eleq}) and (\ref{fdef}), we can write
\begin{equation}\label{e30}
M(s,\hat{\psi}) = - \frac{s\,(V^{(0)}-V_{EB\,y}^{(0)})}{w}\,{\cal L}(\hat{\psi})  + \frac{s\,(V^{(0)}-V)}{x_w}\,{\cal F}(\hat{\psi}),
\end{equation}
and
\begin{equation}\label{e31}
M(s,\hat{\psi})+d_\beta\,\tau\,L(x,\hat{\psi}) = - \frac{s\,(V^{(0)}-V_{i\,y}^{(0)})}{w}\,{\cal L}(\hat{\psi})  + \frac{s\,(V^{(0)}-V)}{x_w}\,{\cal F}(\hat{\psi}).
\end{equation}
Here, $V_{EB\,y}^{(0)}=(V_{i\,y}^{(0)}+\tau\,V_{e\,y}^{(0)})/(1+\tau)$ is the unperturbed ${\bfm E}\times{\bfm B}$
velocity, $V_{i\,y}^{(0)}$ the unperturbed ion velocity, and
$V_{e\,y}^{(0)}$ the unperturbed electron velocity.
[Note that $V_{\ast\,y}^{(0)}=V_{i\,y}^{(0)}-V_{e\,y}^{(0)}$.] Furthermore,
$V^{(0)} = (\mu_i\,V_{i\,y}^{(0)} + \mu_e\,V_{e\,y}^{(0)})/(\mu_i
+\mu_e)$ (see Ref.~\onlinecite{island1}) is the unperturbed island phase-velocity, and $V$ the actual  phase-velocity. All of these
velocities are measured in the {\em lab.\ frame}.
Finally, both ${\cal L}(\hat{\psi})$ and ${\cal F}(\hat{\psi})$ are zero for 
$\hat{\psi}<1$, whereas 
\begin{equation}
{\cal L}(\hat{\psi})=\frac{1}{\langle\!\langle X^2\rangle\!\rangle},
\end{equation}
and
\begin{equation}\label{fsdef}
{\cal F}(\hat{\psi})=
\left.\int_{1}^{\hat{\psi}}\frac{d\hat{\psi}}{\langle\!\langle
X^4\rangle\!\rangle}\right/ \int_{1}^{\infty}\frac{d\hat{\psi}}{\langle\!\langle
X^4\rangle\!\rangle}
\end{equation}
in the region $\hat{\psi}\geq 1$.

Now
\begin{equation}\label{e34}
{\mit \Delta}'(V) = \frac{4}{w}\,\int_{-1}^\infty
\langle\!\langle \delta J_c\,\cos\theta\rangle\!\rangle\,d\hat{\psi}
\end{equation}
(see Ref.~\onlinecite{rf1}), where ${\mit \Delta}'(V)$,  which is specified
in Eq.~(\ref{del}), is the effective tearing stability index in the presence of the resistive wall. Hence, it follows from Eqs.~(\ref{e29}), (\ref{e30}), (\ref{e31}),
and (\ref{e34}) that
\begin{eqnarray}\label{ruth}
\frac{I_1}{\eta}\,\frac{d W}{dt} &=& {\mit\Delta}'(V)
+I_2\,\frac{(V^{(0)}-V_{EB\,y}^{(0)})\,(V^{(0)}-V_{i\,y}^{(0)})}{(W/4)^3}\nonumber\\[0.5ex]&&-I_3\,\frac{2\,(V^{(0)}-[V_{EB\,y}^{(0)}+V_{i\,y}^{(0)}]/2)\,
(V^{(0)}-V)}{x_w\,(W/4)^2}+I_4\,\frac{(V^{(0)}-V)^2}{x_w^{\,2}\,(W/4)},
\end{eqnarray}
where
\begin{eqnarray}
I_1&=&2\int_{-1}^\infty\frac{\langle\!\langle\cos\theta\rangle\!\rangle^2}{
\langle\!\langle 1\rangle\!\rangle}\,d\hat{\psi}=0.823,\\[0.5ex]
I_2 &=& \int_1^\infty\left(
\langle\!\langle X^4\rangle\!\rangle-\frac{\langle\!\langle
X^2\rangle\!\rangle^2}{\langle\!\langle 1\rangle\!\rangle}\right)
\frac{d({\cal L}^2)}{d\hat{\psi}}\,d\hat{\psi}=1.38,\\[0.5ex]
I_3&=&\int_1^\infty\left(
\langle\!\langle X^4\rangle\!\rangle-\frac{\langle\!\langle
X^2\rangle\!\rangle^2}{\langle\!\langle 1\rangle\!\rangle}\right)
\frac{d({\cal L}\,{\cal F})}{d\hat{\psi}}\,d\hat{\psi}=0.195,\\[0.5ex]
I_4&=&\int_1^\infty\left(
\langle\!\langle X^4\rangle\!\rangle-\frac{\langle\!\langle
X^2\rangle\!\rangle^2}{\langle\!\langle 1\rangle\!\rangle}\right)
\frac{d({\cal F}^2)}{d\hat{\psi}}\,d\hat{\psi}=0.469.
\end{eqnarray}

Equation~(\ref{ruth}) is the Rutherford island width evolution
equation\cite{ruth} for a propagating magnetic island interacting with a resistive wall. There are
{\em three} separate ion polarization terms on the right-hand side
of this equation. The first (second term on r.h.s.)\ is the drift-MHD polarization
term for an isolated island (see Ref.~\onlinecite{island1}), and is unaffected by wall braking. This
term, which varies as $W^{-3}$, is stabilizing provided that the unperturbed island phase-velocity
lies between the unperturbed ion fluid velocity and the unperturbed
${\bfm E}\times {\bfm B}$ velocity, and is destabilizing otherwise. The third (fourth term on r.h.s.)\ is the single-fluid MHD polarization term due to
the island velocity-shift induced by wall braking (see Ref.~\onlinecite{wf}).
This term is {\em always destabilizing}, and varies as
$W^{-1}$ and the square of the wall-induced
velocity-shift. The second (third term on r.h.s.)\
is a hybrid of the other two polarization terms. The
sign of this term depends on many factors. However, in the
limit of small electron viscosity (compared to the ion viscosity), when
the unperturbed island phase-velocity lies close to the unperturbed velocity
of the ion fluid,\cite{island1} the hybrid term is stabilizing provided $V_{\ast\,y}^{(0)}\,V^{(0)}>0$, and destabilizing otherwise. In other words,
the hybrid term is {\em stabilizing}\/ if the unperturbed island propagates in the
{\em ion diamagnetic direction} (in the lab.\ frame), and {\em destabilizing}\/ if it propagates in the {\em electron diamagnetic
direction}. The hybrid polarization term varies as $W^{-2}$, and is directly proportional to the
wall-induced island velocity-shift.

\section{Interaction with a static  external resonant magnetic perturbation}
\subsection{Introduction}
Let the walls bounding the plasma at $x=\pm x_w$
now be non-conducting ({\em i.e.}, $\tau_w\rightarrow 0$). Suppose that an even (in $x$) static magnetic perturbation (with the same wave-length
as the magnetic island in the plasma) is generated by currents flowing in field-coils
located in the vacuum region beyond the walls. 

The no-wall tearing stability
index, ${\mit\Delta}_{nw}$, is defined in Sect.~\ref{rint}. The coil eigenfunction, $\psi_c(x)$,
is the continuous even solution to Eq.~(\ref{tear}) which satisfies $\psi_c(0)=0$
and $\psi_c(\pm x_w)=1$. In general, this eigenfunction has
a gradient discontinuity at $x=0$. It is helpful to define
${\mit\Delta}_c=[d\psi_c/dx]_{0-}^{0+}$.

According to standard analysis,\cite{rf} the effective
tearing stability index, ${\mit\Delta}'=[d\ln\psi/dx]_{0-}^{0+}$,
in the presence of an external magnetic perturbation is
\begin{equation}\label{del2}
{\mit\Delta}'(t) = {\mit\Delta}_{nw} + {\mit\Delta}_c\,\frac{{\mit\Psi}_c}
{\mit\Psi}\,\cos\varphi(t),
\end{equation}
where ${\mit\Psi}(t)=|\psi^{(1)}(0,t)|$ is the reconnected magnetic
flux, which is assumed to vary slowly in time,
and ${\mit\Psi}_c$ the flux at the walls solely due to
currents flowing in the external coils. Furthermore, $\varphi(t)$ is
the phase of the island measured with respect to that of the external magnetic
perturbation. Since the external perturbation is stationary, it follows
that
\begin{equation}\label{eom1}
\frac{d\varphi}{dt} = k\,V(t),
\end{equation}
where $V(t)$ is the instantaneous island phase-velocity. Also, the net $y$-directed
electromagnetic force acting on the island takes the form
\begin{equation}\label{lock}
f_y(t) = -\frac{k}{2}\,{\mit\Delta}_c\,{\mit\Psi}\,{\mit\Psi}_c\,\sin\varphi(t).
\end{equation}
Note that, unlike the braking force due to a resistive wall, this
force {\em oscillates} in sign as the island propagates.

\subsection{Determination of flow profiles}
We can reuse the analysis of Sect.~\ref{flow}, except that we must
allow for {\em time dependence} of the function $F$ to take into account the
{\em oscillating} nature of the locking force exerted on the island by the external perturbation. Hence, we write
\begin{equation}
M(s,\psi,t) =  - \frac{d_\beta\,(\mu_i\,\tau-\mu_e)}
{(\mu_i+\mu_e)}\,L(s,\psi) + F(s,\psi,t),
\end{equation}
where
\begin{equation}
L(s,\hat{\psi}) = \frac{s\,V_{\ast\,y}^{(0)}}{w\,d_\beta\,(1+\tau)}\,
\frac{1}{\langle\!\langle X^2\rangle\!\rangle},
\end{equation}
and
\begin{equation}\label{feq}
\frac{\partial}{\partial\psi}\!\left[(\mu_i+\mu_e)\,\frac{\partial}{\partial\psi}\!\left(\langle x^4\rangle\,\frac{\partial F}{\partial\psi}\right) - \langle x^2\rangle\,\frac{\partial
F}{\partial t}\right]=0.
\end{equation}

In order to proceed further, we adopt the separable form approach to
solving Eq.~(\ref{feq}) which was introduced and justified in Ref.~\onlinecite{rf1}.
In other words, we try the following solution:
\begin{equation}\label{ff}
F(s,\psi,t)=s\,F_1(\psi)\,\sin\left(\int_0^t\,k\,V(t')\,dt'\right)
+ s\,F_2(\psi)\,\cos\left(\int_0^t\,k\,V(t')\,dt'\right).
\end{equation}
Of course, $F_1(\psi)$ and $F_2(\psi)$ are both zero within the island
separatrix. Furthermore,
\begin{eqnarray}\label{bc1}
|x|\, F_1&\rightarrow& F_0,\\[0.5ex]
|x|\,F_2&\rightarrow&0,\label{bc2}
\end{eqnarray}
as $|x|/w\rightarrow \infty$. Here, $F_0$ is a constant. The above boundary conditions imply that the
function $F(s,\psi,t)$ corresponds to a velocity profile which is
{\em localized} in the vicinity of the island.

Matching to the outer region yields
\begin{equation}\label{ff1}
F_0\,\sin\left(\int_0^t\,k\,V(t')\,dt'\right) = V^{(0)}-V(t).
\end{equation}
Hence, differentiating with respect to $t$, we obtain
\begin{equation}\label{ff2}
\frac{1}{k\,V}\,\frac{dV}{dt} = -\,F_0\,\cos\left(\int_0^t\,k\,V(t')\,dt'\right), 
\end{equation}
and
\begin{equation}\label{ff3}
\frac{d}{dt}\!\left(\frac{1}{k\,V}\,\frac{dV}{dt}\right) = k\,V\,(V^{(0)}-V).
\end{equation}

Substituting Eq.~(\ref{ff}) into Eq.~(\ref{feq}), and integrating
once in $\psi$ using the boundary conditions (\ref{bc1}) and (\ref{bc2}),
we get
\begin{eqnarray}\label{ef1}
{\rm sgn}(V)\,\frac{\lambda^2}{2\,w^2}\,\frac{d}{d\hat{\psi}}\!\left(\langle\!\langle X^4\rangle\!\rangle\,
\frac{d F_1}{d\hat{\psi}}\right) + \langle\!\langle X^2\rangle\!\rangle\,F_2
&=&0,\\[0.5ex]
{\rm sgn}(V)\,\frac{\lambda^2}{2\,w^2}\,\frac{d}{d\hat{\psi}}\!\left(\langle\!\langle X^4\rangle\!\rangle\,
\frac{d F_2}{d\hat{\psi}}\right) - \langle\!\langle X^2\rangle\!\rangle\,F_1
&=&-\frac{F_0}{w}.\label{ef2}
\end{eqnarray}
Here, $\lambda = \sqrt{2\,(\mu_i+\mu_e)/k\, |V|}$ is the localization
scale-length of the velocity profile corresponding to the function $F$.

Suppose that $w\ll \lambda \ll x_w$. In other words, suppose
that the localization scale-length of the velocity profile associated with
$F$ is
much larger than the island width, but much smaller than the extent
of the plasma. In this limit (which corresponds to the ``weakly
localized'' regime of Ref.~\onlinecite{rf1}),  Eqs.~(\ref{ef1})
and (\ref{ef2}) can be solved to give
\begin{eqnarray}
|X|\,F_1 &=& \frac{F_0}{w}\left[1-
\exp\!\left(-\frac{w\,|X|}{\lambda}\right)
\,\cos\!\left(\frac{w\,|X|}{\lambda}\right)\right]\,{\cal F}(\hat{\psi}),\\[0.5ex]
|X|\,F_2 &=& {\rm sgn}(V)\,\frac{F_0}{w}\,\exp\!\left(-\frac{w\,|X|}{\lambda}\right)
\,\sin\!\left(\frac{w\,|X|}{\lambda}\right)\,{\cal F}(\hat{\psi}).
\end{eqnarray}
Here, ${\cal F}(\hat{\psi})$ is specified in
Eq.~(\ref{fsdef}).
It follows from Eqs.~(\ref{ff}), (\ref{ff1}), and (\ref{ff2}) that
\begin{eqnarray}\label{fnew}
F(s,\hat{\psi},t) &=& \frac{s}{w}\,(V^{(0)}-V)\left[1-
\exp\!\left(-\frac{w\,|X|}{\lambda}\right)
\,\cos\!\left(\frac{w\,|X|}{\lambda}\right)\right]\,\frac{{\cal F}(\hat{\psi})}{|X|}
\nonumber\\[0.5ex]
&&-\frac{s}{w}\,\frac{1}{k\,|V|}\,\frac{dV}{dt}\,\exp\!\left(-\frac{w\,|X|}{\lambda}\right)
\,\sin\!\left(\frac{w\,|X|}{\lambda}\right)\,\frac{{\cal F}(\hat{\psi})}{|X|}.
\end{eqnarray}

\subsection{Island equation of motion}\label{eomx}
Reusing the analysis of Sect.~\ref{fbal},
taking into account the time dependence of $F$, we obtain
\begin{equation}
f_y= 2\,s\,(\mu_i+\mu_e)\lim_{x/w\rightarrow \infty}
\left[x^2\,\frac{\partial}{\partial x}\!\left(
\frac{1}{x}\,\frac{\partial(x\,F)}{\partial x}\right)\right] -2\,\frac{\partial}{\partial t}\!\int_{-{\mit\Psi}}^{-\infty}
\left(\langle x^3\rangle\,\frac{\partial F}{\partial\psi}-\langle
x\rangle\,F\right)\,d\psi.
\end{equation}
According to the boundary conditions (\ref{bc1}) and (\ref{bc2}),
the first term on the right-hand side is identically zero. Transforming the
second term on the right-hand side, using the fact that the integral
is dominated by the region $|X|\gg 1$, we get
\begin{equation}
f_y = -2\,s\,{\mit\Psi}\,\frac{\partial}{\partial t}\!\int_0^\infty
X\,\frac{\partial(X\,F)}{\partial X}\,dX.
\end{equation}
Finally, Eqs.~(\ref{ff2}), (\ref{ff3}), and (\ref{fnew}) yield
\begin{equation}\label{fydef}
f_y = \lambda\left[\frac{d V}{dt} + k\,|V|(V-V^{(0)})\right].
\end{equation}

Making use of Eq.~(\ref{lock}),  the island equation of motion takes the form:
\begin{equation}\label{eom}
\sqrt{\frac{2\,(\mu_i+\mu_e)}{k\,|V|}}\,\frac{dV}{dt}
+ \sqrt{2\,(\mu_i+\mu_e)\,k\,|V|}\,(V-V^{(0)})+
\frac{k}{2}\left(\frac{W}{4}\right)^2\left(\frac{W_c}{4}\right)^2\,\sin\varphi=0.
\end{equation}
Here, $(W_c/4)^2={\mit\Delta}_c\,{\mit\Psi}_c$.
The first term on the left-hand side represents the inertia
of the region of the plasma (of width $\sqrt{2\,(\mu_i+\mu_e)/k\,|V}|$) which is viscously coupled to the island,
the second term represents the viscous restoring force, and the third 
term represents the
locking force due to the external perturbation. Note that the above
equation is identical to that obtain from single-fluid MHD theory.\cite{rf1}
The above analysis is valid provided $w\ll \sqrt{2\,(\mu_i+\mu_e)/k\,|V|}\ll x_w$.

\subsection{Determination of ion polarization correction}\label{polz1}
Reusing the analysis of Sect.~\ref{polz}, we obtain
\begin{equation}
\delta J_c = -\frac{1}{2}\!\left(X^2-\frac{\langle\!\langle X^2\rangle\!\rangle}{\langle\!\langle 1
\rangle\!\rangle}\right) \frac{\partial}{\partial\hat{\psi}}\!\left[M\,(M+d_\beta\,\tau\,L)\right]
+\eta^{-1}\,\frac{d{\mit\Psi}}{dt}\,\frac{\langle\!\langle\cos\theta\rangle\!\rangle}{\langle\!\langle
1\rangle\!\rangle},
\end{equation}
where
\begin{equation}
M(s,\hat{\psi},t) = - \frac{s\,(V^{(0)}-V_{EB\,y}^{(0)})}{w}\,{\cal L}(\hat{\psi})  - \frac{s\,f_y(t)}{2\,(\mu_i+\mu_e)}\,{\cal F}(\hat{\psi}),
\end{equation}
and
\begin{equation}
M(s,\hat{\psi},t)+d_\beta\,\tau\,L(x,\hat{\psi}) = - \frac{s\,(V^{(0)}-V_{i\,y}^{(0)})}{w}\,{\cal L}(\hat{\psi}) -\frac{s\,f_y(t)}{2\,(\mu_i+\mu_e)}\,{\cal F}(\hat{\psi}).
 \end{equation}
 Here, use has been made of Eqs.~(\ref{fnew}) and (\ref{fydef}),
 as well as the fact that the polarization term integral is dominated
 by the region $|X|\sim O(1)$. Finally, Eqs.~(\ref{e34}),
 (\ref{del2}), and (\ref{lock}) yield
 \begin{eqnarray}
 \frac{I_1}{\eta}\,\frac{dW}{dt}&=&{\mit\Delta}_{nw} + 
 \left(\frac{W_c}{W}\right)^2\,\cos\varphi
 + I_2\,\frac{(V^{(0)}-V_{EB\,y}^{(0)})\,(V^{(0)}-V_{i\,y}^{(0)})}{(W/4)^3}\nonumber\\[0.5ex]
 &&- I_3\,\frac{k}{2}\,\frac{(V^{(0)}-[V_{EB}^{(0)}+V_{i\,y}^{(0)}]/2)}{(\mu_i+\mu_e)}\,
 \left(\frac{W_c}{4}\right)^2\sin\varphi\nonumber\\[0.5ex]
&& + I_4\,\frac{k^2}{16\,(\mu_i+\mu_e)^2}\left(\frac{W}{4}\right)^3\left(
 \frac{W_c}{4}\right)^4\,\sin^2\varphi,\label{eggg}
 \end{eqnarray}
 where $I_1$, $I_2$, $I_3$, and $I_4$ are specified in Sect.~\ref{polz}.
 
 Equation~(\ref{eggg}) is the Rutherford island width evolution equation
 for a propagating island interacting with a static  external resonant magnetic perturbation.
 There are
{\em three} separate ion polarization terms on the right-hand side
of this equation. The first (third term on r.h.s.)\ is the drift-MHD polarization
term for an isolated island (see Ref.~\onlinecite{island1}), and is unaffected by the external perturbation.
The third (fifth term on r.h.s.)\ is the single-fluid MHD polarization term due to the
oscillation in island phase-velocity induced by the external perturbation (see
Ref.~\onlinecite{rf1}).
This term modulates as the island propagates, but is {\em always destabilizing}. The second (fourth term on r.h.s.)\ is a
hybrid of the other two polarization terms.

\subsection{Solution of island equations of motion}\label{sveom}
Let us solve  the island equations of motion, (\ref{eom1}) and (\ref{eom}),
in the limit in which the external magnetic perturbation is sufficiently weak that
it does not significantly perturb the island phase-velocity. 
Let us  also assume that $\eta$ is
so small  that the island width, $W$, does not vary
appreciably   with island phase.
In this limit, we can write
\begin{equation}\label{eav}
\varphi(t)= k\,V^{(0)}\,t  + \alpha_s\,\sin(k\,V^{(0)}\,t) + \alpha_c\,
\cos(k\,V^{(0)}\,t),
\end{equation}
where $|\alpha_s|, |\alpha_c|\ll 1$. Substitution of the above expression
into Eqs.~(\ref{eom1}) and (\ref{eom}) yields
\begin{equation}
\alpha_s \simeq \left.\left(\frac{W}{4}\right)^2\left(\frac{W_c}{4}\right)^2\right/
4\,\lambda\,[V^{(0)}]^2,
\end{equation}
and $\alpha_c \simeq {\rm sgn}(V^{(0)})\,\alpha_s$,
where $\lambda = \sqrt{2\,(\mu_i+\mu_e)/k\,|V^{(0)}|}$ is the velocity
localization scale-length.
Averaging over island phase, using Eq.~(\ref{eav}), we obtain
\begin{eqnarray}
\overline{\cos\varphi}&\simeq& - \frac{\alpha_s}{2},\\[0.5ex]
\overline{\sin\varphi} &\simeq& {\rm sgn}(V^{(0)})\,\frac{\alpha_s}{2},\\[0.5ex]
\overline{\sin^2\varphi}&\simeq&\frac{1}{2}.
\end{eqnarray}
Hence, the average of the Rutherford island width evolution equation,
(\ref{eggg}), over island phase takes the form
\begin{eqnarray}
\frac{I_1}{\eta}\,\frac{dW}{dt}&=&{\mit\Delta}_{nw} + 
   I_2\,\frac{(V^{(0)}-V_{EB\,y}^{(0)})\,(V^{(0)}-V_{i\,y}^{(0)})}{(W/4)^3}\nonumber\\[0.5ex]
   &&-\frac{\alpha_s}{2}\left(\frac{W_c}{W}\right)^2\left\{
   1 +I_3\,\frac{(V^{(0)}-[V_{EB}^{(0)}+V_{i\,y}^{(0)}]/2)}{V^{(0)}}
   \left(\frac{w}{\lambda}\right)^2- I_4\left(\frac{w}{\lambda}\right)^3
   \right\}.\label{e70}
\end{eqnarray}
The first two terms on the right-hand side of the above
equation are the intrinsic tearing mode drive and the drift-MHD
polarization term, respectively, and are unaffected by the external perturbation. 
The next three terms (within the curly braces) are the {\em phase-averaged} external perturbation drive, hybrid
polarization term, and single-fluid MHD polarization term, respectively.
It can be seen that the external perturbation drive is on average {\em stabilizing},
whereas the single-fluid MHD polarization term is {\em
destabilizing}.\cite{rf} The
sign of the hybrid term depends on many factors. However, in the
limit of small electron viscosity (compared to the ion viscosity), when
the unperturbed island phase-velocity lies close to the unperturbed velocity
of the ion fluid,\cite{island1} the hybrid term is on average stabilizing provided $V_{\ast\,y}^{(0)}\,V^{(0)}>0$, and destabilizing otherwise. In other words,
the hybrid term is {\em stabilizing}\/ if the unperturbed island  propagates in the
{\em ion diamagnetic direction} (in the lab.\ frame), and {\em destabilizing}\/ if it  propagates in the {\em electron diamagnetic
direction}. 
Finally, since our analysis is based on the fairly reasonable assumption that $w/\lambda\ll 1$,
it follows from Eq.~(\ref{e70}) that the phase-averaged external perturbation drive dominates the
phase-averaged hybrid and single-fluid MHD polarization terms. Hence, we conclude that, on average,
an island propagating in  the presence of an external magnetic perturbation
experiences a net {\em stabilizing} effect. 

\section{Summary and discussion}
We have investigated the dynamics of a propagating magnetic island
interacting with a resistive wall or a static external resonant magnetic perturbation
using {\em two-fluid}, drift-MHD theory in {\em slab geometry}. In both cases, we
find that the island equation of motion takes exactly the same form
as that predicted by single-fluid MHD theory (see Sects.~\ref{fbal} and
\ref{eomx}). However, two-fluid
effects do give rise to additional ion polarization terms in the 
Rutherford island width evolution equation. 

In general, we find
that there are {\em three} separate ion polarization terms in the Rutherford
equation (see Sects.~\ref{polz} and
\ref{polz1}). The first is the drift-MHD polarization term for an isolated
island, and is completely unaffected by interaction with a
resistive wall or an external magnetic perturbation. Next, there is the
polarization term due to interaction with a resistive wall or magnetic
perturbation which is predicted by single-fluid MHD theory. This term is
always {\em destabilizing}. Finally, there is a hybrid of the other
two polarization terms. The
sign of this term depends on many factors. However, in the
limit of small electron viscosity (compared to the ion viscosity), when
the unperturbed island phase-velocity lies close to the unperturbed velocity
of the ion fluid,\cite{island1} the hybrid term is stabilizing if the unperturbed island  propagates in the
{\em ion diamagnetic direction} (in the
lab.\ frame), and {\em destabilizing}\/ if it propagates in the {\em electron diamagnetic
direction}.

It is also demonstrated that a propagating magnetic island interacting
with a static external resonant magnetic perturbation generally experiences a net
{\em stabilizing} effect (see Sect.~\ref{sveom}). This
follows because in the Rutherford island width evolution equation
the phase-averaged 
 drive term due to the external perturbation (which is stabilizing)  is generally much larger than either the phase-averaged hybrid polarization term (which can be destabilizing) or the phase-averaged
single-fluid MHD polarization term (which is destabilizing).

\subsection*{Acknowledgments}
This research was  funded by
the U.S.\ Department of Energy under contract DE-FG05-96ER-54346.

\end{document}